\documentclass[aps,prb,twocolumn,showpacs,twoside,amsmath, amsfonts]{revtex4}
\usepackage{psfrag}
\usepackage{graphicx}
\usepackage{epsfig}
\usepackage{amssymb}

\begin{document}
\title{One-dimensional Anderson Localization: Devil's Staircase of Statistical Anomalies.}
\author{V.E.Kravtsov$^{1,2}$ and V.I.Yudson$^{3}$}
\affiliation{$^{1}$The Abdus Salam International Centre for
Theoretical Physics, P.O.B. 586, 34100 Trieste, Italy.\\
$^{2}$Landau Institute for Theoretical Physics, 2 Kosygina
st.,117940 Moscow, Russia.\\$^{3}$Institute for Spectroscopy,
Russian Academy of Sciences, 142190 Troitsk, Moscow reg., Russia.}

\date{today}
\begin{abstract}
The statistics of wavefunctions in the one-dimensional (1d) Anderson
model of localization is considered. It is shown that at any energy
that corresponds to a rational filling factor $f=\frac{p}{q}$ there
is a statistical anomaly which is seen in expansion of the
generating function (GF) to the order $q-2$ in the disorder
parameter. We study in detail the principle anomaly at
$f=\frac{1}{2}$ that appears in the leading order. The
transfer-matrix equation of the Fokker-Planck type with a
two-dimensional internal space is derived for GF. It is shown that
the zero-mode variant of this equation is integrable and a solution
for the generating function is found in the thermodynamic limit.
\end{abstract}
 \maketitle

--{\it Introduction.} Anderson localization (AL) enjoys an unusual
fate of being a subject of advanced research during a half of
century. The seminal paper by P.W.Anderson \cite{Anderson} opened up
a direction of research on the interplay of quantum mechanics and
disorder which is of fundamental interest up to now
\cite{Mirlin2008}. The one-dimensional tight-binding model with
diagonal disorder --the Anderson model (AM)-- which is the simplest
and the most studied model of this type, became a paradigm of AL:
\begin{equation}
\label{Ham}
H=\sum_{i}\varepsilon_{i}\,c^{\dagger}_{i}c_{i}-\sum_{i}t_{i}\left(\,c^{\dagger}_{i}c_{i+1}+
c^{\dagger}_{i+1}c_{i}\right).
\end{equation}
In this model the hopping integral is deterministic $t_{i}=t$ and
the on-site energy $\varepsilon_{i}$ is a random Gaussian variable
uncorrelated at different sites and characterized by the variance
$\langle(\delta\varepsilon_{i})^{2}\rangle=w$.
The dimensionless parameter $\alpha^{2}=w/t^{2}$ describes the
strength of disorder.

The best studied is the continuous limit of this model in which the
lattice constant $a\rightarrow 0$ at $ta^{2}$ remaining finite
\cite{Ber, AR, Mel, Kolok}. There was also a great deal of activity
\cite{1dRev, Pastur} aimed at a rigorous mathematical description of
1d AL. However, despite considerable efforts invested, some subtle
issues concerning 1d AM still remain unsolved. One of them is the
effects of commensurability between the de-Broglie wavelength
$\lambda_{E}$ (which depends on the energy $E$) and the lattice
constant $a$. The parameter that controls the commensurability
effects is the filling factor $f=\frac{2a}{\lambda_{E}}$ (fraction
of states below the energy $E$).

It was known for quite a while \cite{Derrida} that the Lyapunov
exponent (which is essentially the inverse localization length)
takes anomalous values at the filling factors equal to $\frac{1}{2}$
and $\frac{1}{3}$ (compared to those at filling factors $f$ beyond
the window of the size $\alpha^{2}\ll 1$ around $f=\frac{1}{2}$ and
$f=\frac{1}{3}$, see Fig.1). Recently \cite{Titov, AL} it was found
that the statistics of conductance in 1d AM is anomalous at the
center of the band that corresponds to the filling factor
$f=\frac{1}{2}$. We want to stress that all these anomalies were
observed for the AM Eq.(\ref{Ham}) in which the on-site energy
$\varepsilon_{i}$ is random. This Hamiltonian does not possess the
{\it chiral symmetry} \cite{Dyson, Mirlin2008} which is behind the
statistical anomalies at the center of the band $E=0$ in the {\it
Lifshitz model} described by Eq.(\ref{Ham}) with the deterministic
$\varepsilon_{i}=0$ and a random hopping integral $t_{i}$. Thus the
statistical anomalies in the 1d AM raise a question about {\it
hidden symmetries} which do not merely reduce to the two-sublattice
division \cite{Dyson, AL, Mirlin2008}.
\begin{figure}
\label{fig1}
\includegraphics[width=6cm, height=5cm]{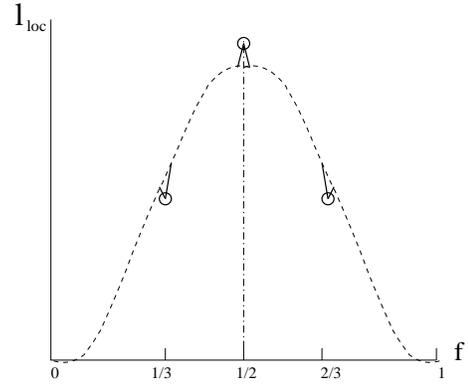}
\caption{Schematic representation of statistical anomalies in the
localization radius. The dashed line represents the "bare"
localization radius $\ell_{0}=a\frac{2t^{2}}{w}\,\sin^{2}(\pi f)$;
the circles give the localization length at a filling factor being a
simple fraction and solid lines give a sketch of  behavior in the
perturbed window.}
\end{figure}
Finally, the {\it sign} of the anomaly is different for the center
of the band $f=\frac{1}{2}$ and the filling factor $f=\frac{1}{3}$.
All these observations point out to a new phenomenon of the devil's
staircase type which is essentially due to {\it disorder} and its
interplay with the Bragg scattering off the underlying lattice.

In this Letter we study the {\it generating function} (GF)
$\Phi(u,\phi;x)$ that allows to compute {\it all} local statistical
properties of 1d AM. The simplest of them is the statistics of the
wavefunction amplitude  $|\psi(x)|^{2}$ characterized by the moments
$I_{m}=\langle |\psi(x)|^{2m}\rangle\,\ell_{0}^{m}$:
\begin{eqnarray}
\label{moment1}
I_{m}&=&\frac{2}{(m-2)!}\,\int_{0}^{\pi}\frac{d\phi}{\pi}\,
\cos^{2m}\phi\,\\
\nonumber&\times&\int_{0}^{\infty}du\,u^{m-2}\,\Phi(u,\phi;x)\,
\Phi(u,-\phi-2\pi f;L-x),
\end{eqnarray}
where $L$ is the total length of the system and
$\ell_{0}=a\frac{2t^{2}}{w}\,\sin^{2}(\pi f)$ is the "bare"
localization length.

We  derive the corresponding transfer-matrix equation (TME) for the
GF
\begin{equation}
\label{TME}
\partial_{x} \Phi(u,\phi;x)=[\hat{L}_{f}(u,\phi)-u]\,\Phi(u,\phi;x),
\end{equation}
which {\it zero-mode} variant ($\Phi(u,\phi;x)\equiv \Phi(u,\phi)$
is independent of the space coordinate $x$) appears to be a partial
differential equation (PDE) depending on {\it two} variables. One of
them (denoted by $u$)is associated with the amplitude of the
wavefunction $\psi\sim \sqrt{u} \cos\phi$ while the other (denoted
by $\phi$)has a physical meaning of its phase. The well known
\cite{Kolok} TME in the continuous limit $f\ll 1$ can be obtained by
the averaging of this PDE over the phase variable $\phi$ thus
reducing it to an ODE in the {\it single} variable $u$.

We  show that there are statistical anomalies at {\it any} rational
filling factor $f=\frac{p}{q}$. Namely, the operator
$\hat{L}_{f}(u,\phi)$ in Eq.(\ref{TME}) expanded in the disorder
parameter $\alpha^{2}$
\begin{equation}
\label{expan}
\hat{L}_{f}(u,\phi)=\hat{L}_{f}^{(0)}(u,\phi)+\alpha^{2}\,\hat{L}_{f}^{(1)}(u,\phi)+
\alpha^{4}\,\hat{L}_{f}^{(2)}(u,\phi)+...
\end{equation}
is such that
\begin{equation}
\label{deltaL} \hat{L}_{f}^{(n)}=\hat{L}_{f}^{(n,{\rm
reg})}+\sum_{p=1}^{n+1}\Delta\hat{L}^{(n)}_{p}\,\delta\left(f,\frac{p}{n+2}\right)
\end{equation}
contains a regular part $\hat{L}_{f}^{(n,{\rm reg})}$ with a smooth
dependence on $f$ and an anomalous part that appears only at
$f=\frac{1}{2+n},\frac{2}{2+n},...\frac{n+1}{2+n}$. In the leading
order $(n=0)$ in $\alpha^{2}$ the anomalous term appears only at
$f=\frac{1}{2}$. In the next order one can observe anomalies at
$f=\frac{1}{3}$ and $\frac{2}{3}$, etc. Though anomalous terms
corresponding to the denominator $q>2$ are small at weak disorder,
they have an abrupt dependence on $f$. This allows to speak about
the "devil's staircase of anomalies".

We study in detail the principal anomaly at $f=\frac{1}{2}$.
Remarkably, the corresponding zero-mode TME appears to be {\it
integrable}. We find a unique solution to this equation which
describes any local statistics of wavefunctions in the center of the
band.

-- {\it Derivation of the TM equation.} The starting point of our
analysis is the TM equation for the generating function
$\Phi_{j}(u,\phi)$ on the lattice site $j$:
\begin{equation}
\label{op}
\Phi_{j+1}(u,\phi)=\left(1+\frac{2a}{\ell_{0}}\,\left[{\cal L
}(u,\phi)-c_{1}(\phi)\,u\right]\right)\,\Phi_{j}(u,\phi-\pi f),
\end{equation}
where ${\cal
L}(u,\phi)=c_{2}(\phi)\,u^{2}\partial^{2}_{u}+c_{3}(\phi)\,(u\partial_{u}-1)+c_{4}(\phi)\,
u\partial_{u}\partial_{\phi}+c_{5}(\phi)\,\partial_{\phi}+c_{6}(\phi)\,\partial^{2}_{\phi}$.
The coefficients $c_{i}(\phi)$ are all combinations of $\cos(2\phi)$
and $\sin(2\phi)$ which at first glance do not show any nice
structure: $c_{1}(\phi)=\frac{1}{2}(1+\cos(2\phi))$,
$c_{2}(\phi)=1-\cos^{2}(2\phi)$,
$c_{3}(\phi)=-(1-\cos(2\phi)-2\cos^{2}(2\phi))$,
$c_{4}(\phi)=\sin(2\phi)(1+\cos(2\phi))$,
$c_{5}(\phi)=-\frac{3}{2}\sin(2\phi)(1+\cos(2\phi))$,
$c_{6}(\phi)=\frac{1}{4}(1+\cos(2\phi))^{2}$.

This equation has been derived in Ref.\cite{OsK} by expansion to the
first order in $\alpha^{2}$ of the exact integral TM equation
obtained by the super-symmetry method \cite{Efet-book}. By
construction \cite{OsK} the function $\Phi_{j}(u,\phi)$ must be
periodic in $\phi$ with the period of $\pi$ which corresponds to the
phase factor $\cos\phi$ of the wave function sweeping all possible
values in the interval $[0,\pi]$. However, the shift in the argument
$\phi$ in the r.h.s. of Eq.(\ref{op}) is by a {\it fraction} $f$ of
$\pi$. For a rational $f=\frac{p}{q}$ one has to make $q$ iterations
in Eq.(\ref{op}) in order to get a closed equation for the GF. In
the leading order in $\alpha$ we obtain:
\begin{eqnarray}
\label{lin} &&\Phi_{j+q}(u,\phi)-\Phi_{j}(u,\phi)
=\frac{2a}{\ell_{0}}\\ &\times&\left[\sum_{r=0}^{q-1}{\cal
L}(\phi-r\,\pi p/q)-u\sum_{r=0}^{q-1}c_{1}(\phi-r\,\pi p/q)
\right]\,\Phi_{j}(u,\phi).\nonumber
\end{eqnarray}
The reason for the anomaly is the following identity that shows a
jump at $q=2$:
\begin{equation}
\label{id1} \sum_{r=0}^{q-1}e^{2i\phi-2i r\,\pi p/q}=0,\;\;
\sum_{r=0}^{q-1}e^{4i\phi-4i r\,\pi p/q }=\left\{\begin{matrix}0, &
q>2\cr 2e^{4i\phi},& q=2\cr
\end{matrix} \right.
\end{equation}
One can see from this identity that for $q>2$ the summation in
Eq.(\ref{lin}) is the same as averaging over $\phi$: all the
$\phi$-dependent terms vanish in both cases. Assuming $q\ll
\ell_{0}/a$, expanding the l.h.s. of Eq.(\ref{lin}) and introducing
the dimensionless  coordinate $x=ja/\ell_{0}$ we obtain:
\begin{equation}
\label{ord}
\partial_{x}\Phi=\hat{L}_{f}^{(0,{\rm reg})}\Phi=\left[u^{2}\partial^{2}_{u}-u+\frac{3}{4}\partial^{2}_{\phi} \right]
\,\Phi.
\end{equation}
This equation admits the independent of $\phi$ {\it stationary}
solution:
\begin{equation}
\label{stat} \Phi(u,\phi)=e^{-\epsilon
x}\,\sqrt{u}\,K_{\sqrt{1-4\epsilon}}(2\sqrt{u}).
\end{equation}
This solution has been earlier obtained \cite{Kolok} in the
continuous limit $f\ll 1$. It also arises in the theory of a
multi-channel disordered wire \cite{Efet-book, Mirlin2000}. For a
system of the size $L\rightarrow\infty$ only zero mode solution
corresponding to $\epsilon=0$ is relevant. Substituting
Eq.(\ref{stat}) with $\varepsilon=0$ into Eq.(\ref{moment1}) we
found the following distribution function of the eigenfunction
amplitude in a long {\it strictly} one-dimensional system
(amazingly, this result was not known before):
\begin{equation}
\label{1d-dist} {\cal P}(|\psi|^{2})=\frac{\ell_{0}}{L}\,\frac{{\rm
exp}\left(-|\psi|^{2}\ell_{0}\right)}{|\psi|^{2}}.
\end{equation}
This distribution is valid for $|\psi|^{2}\ell_{0}\gg
e^{-L/\ell_{0}}$ and should be cut off at very small $|\psi|^{2}$ to
ensure normalizability \cite{rem1}.

At $q=2$ (and only at $q=2$ in the leading order in $\alpha$) the
$\phi$-dependence in Eqs.(\ref{id1}),(\ref{lin}) survives and gives
rise to the anomalous term \cite{rem4}:
\begin{eqnarray}
\label{del-2}
\Delta\hat{L}^{(0)}&=&\cos(4\phi)\,\left[-u^{2}\partial^{2}_{u}+2u\partial_{u}
+\frac{1}{4}\partial^{2}_{\phi}-2 \right]\nonumber \\
&+&\sin(4\phi)\,\left[u\partial_{u}\partial_{\phi}-\frac{3}{2}\partial_{\phi}
\right].
\end{eqnarray}
Again, like in Eq.(\ref{op}), there is apparently no nice structure
in Eq.(\ref{del-2}). Moreover, because of the anomalous term the
entire operator $\hat{L}_{\frac{1}{2}}^{(0)}(u,\phi)$ acquires an
explicit $\phi$-dependence and thus the {\it zero-mode} TME becomes
a two-variable second-order PDE which no longer admits a
$\phi$-independent solution. Yet it appears exactly solvable!

--{\it Separation of variables.} The integrability of the zero mode
TME for $f=\frac{1}{2}$ is shown in three steps. The step one is to
introduce new set of variables $u$ and $v=u\cos(2\phi)$ instead of
$(u,\phi)$ and a new function
$\tilde{\Phi}(u,v)=u^{-1}\,\Phi(u,\frac{1}{2}\arccos(v/u))$. In
these variables the stationary TME  $[\hat{L}_{\frac{1}{2}}^{(0,{\rm
reg})}+\Delta\hat{L}^{(0)}]\Phi=-\epsilon \Phi$ takes a very
symmetric form:
\begin{eqnarray}
\label{stat-xy}
[D_{1}^{2}+D_{3}^{2}]\,\tilde{\Phi}=
\frac{u}{2}\,\,\tilde{\Phi}-\epsilon\,\tilde{\Phi},
\end{eqnarray}
where the operators $D_{1}$ and $D_{3}$ belong to the family of
three operators from the representation of the $sl_{2}$ algebra:
\begin{equation}
\label{A-xy} D_{1}=\sqrt{u^{2}-v^{2}}\,\,\partial_{u},\;
D_{2}=u\,\partial_{v}+v\,\partial_{u},\;
D_{3}=-\sqrt{u^{2}-v^{2}}\,\,\partial_{v}
\end{equation}
obeying the commutation relations:
\begin{eqnarray}
\label{algebra} [D_{1},D_{2}]=-D_{3},\;
 [D_{3},D_{1}]=D_{2},\;
 [D_{2},D_{3}]=D_{1}.
\end{eqnarray}
Now it is clear that there is a hidden order in a set of
$\phi$-dependent terms in Eq.(\ref{del-2}) and the way they match
the regular part  $\hat{L}_{f}^{(0,{\rm reg})}$ in r.h.s. of
Eq.(\ref{ord}).

The next step is to transform Eq.(\ref{stat-xy}) to the
Schroedinger-like  equation $
-(\partial_{u}^{2}+\partial_{v}^{2})\,\Psi+U(u,v)\,\Psi=0$ for the
function $\Psi(u,v)=(u^{2}-v^{2})^{\frac{1}{4}}\,\tilde{\Phi}$,
where
\begin{eqnarray}
\label{Schroedinger}
U=-\frac{3}{4}\,\frac{u^{2}+v^{2}}{(u^{2}-v^{2})^{2}}+\frac{1}{2}\,
\frac{u}{u^{2}-v^{2}}-\frac{\epsilon}{u^{2}-v^{2}}.
\end{eqnarray}
Finally we introduce the variables
\begin{equation}
\label{xi-eta} \xi=\frac{u+v}{2}=u\,\cos^{2}\phi,\;\;\;\;
\eta=\frac{u-v}{2}=u\,\sin^{2}\phi.
\end{equation}
It is easy to see that the kinetic energy and the first two terms in
Eq.(\ref{Schroedinger}) become the sum of two identical Hamiltonians
$\hat{H}_{\xi}+\hat{H}_{\eta}$ where $\hat{H}_{\xi}$ is given by:
\begin{equation}
\label{H-1d}
\hat{H}_{\xi}=-\partial_{\xi}^{2}-\frac{3}{16}\,\frac{1}{\xi^{2}}+\frac{1}{4\xi}.
\end{equation}
Thus at $\epsilon=0$ we explicitly separated the variables in
Eq.(\ref{stat-xy}) reducing the two-dimensional PDE to two ODE's of
the Schredinger type
$\hat{H}_{\xi}\varphi_{\lambda}(\xi)=\lambda\varphi_{\lambda}(\xi)$
and
$\hat{H}_{\eta}\varphi_{-\lambda}(\eta)=-\lambda\varphi_{-\lambda}(\eta)$.
Each of this equations reduces to the well known Weber's
differential equation which solution is given in terms of the
hypergeometric functions (Whittaker functions) \cite{GR}.

Albeit the above procedure does not work for $\varepsilon\neq 0$
because of the last term in Eq.(\ref{Schroedinger}), the
integrability of the zero-mode TME is a remarkable fact that allows
to describe anomalous statistics in an infinitely long system.

--{\it Uniqueness of the solution.} The general solution to the
"Schroedinger equation" with $\epsilon=0$ in Eq.(\ref{Schroedinger})
is given by the integral over the parameter $\lambda$:
\begin{equation}
\label{gen} \Psi=\int d\lambda
d\bar{\lambda}\;c(\lambda,\bar{\lambda})\;\varphi_{\lambda}(\xi)\,\varphi_{-\lambda}(\eta),
\end{equation}
where integration is generically over the complex plane of $\lambda$
and $c(\lambda,\bar{\lambda})$ is an arbitrary function \cite{rem2}.
How does this huge degeneracy comply with the intuitive expectation
that the statistics of wavefunctions in an infinite disordered chain
should be unique and independent of the boundary conditions? Below
we show that the natural physical requirements on $\Phi(u,\phi)$
help to determine GF up to a constant factor which can be further
fixed using the wave function normalization $\langle
|\psi|^{2}\rangle=\frac{1}{L}$.

First of all we note that
$F(\lambda;\xi,\eta)=\varphi_{\lambda}(\xi)\,\varphi_{-\lambda}(\eta)$
is a holomorphic function of $\lambda$, i.e. it depends only on
$\lambda=\rho e^{i\sigma}$ but not on $\bar{\lambda}=\rho
e^{-i\sigma}$. The idea is to represent the integral over the
complex plane as an integral over $\rho$ and $\sigma$ and then
rotate the contour of integration  $\rho\rightarrow t e^{-i\sigma}$
so that the dependence on $\sigma$ remains only in
$c(\lambda,\bar{\lambda})$ and in the integration measure but not in
$F(\lambda;\xi,\eta)$. Then performing integration over $\sigma$ one
obtains a new function $C(t)=t\int
d\sigma\,e^{-2i\sigma}\,c(t,te^{-2i\sigma})$ which stands for
$c(\lambda,\bar{\lambda})$ in an expression similar to
Eq.(\ref{gen}) but involving only a one-dimensional {\it contour
integral}. This contour can be further rotated to make the
expression more symmetric.  Thus without loss of generality we write
a solution to the zero-mode TM equation Eq.(\ref{stat-xy}) for
$f=\frac{1}{2}$:
\begin{eqnarray}
\label{canon}
&&\Phi(\xi,\eta)=\frac{\xi+\eta}{(\xi\eta)^{1/4}}\int_{0}^{\infty}d\lambda\,
C(\lambda)
\\
&\times&\left[W_{-\lambda\epsilon,\frac{1}{4}}\,\left(
\frac{\bar{\epsilon}\xi}{4\lambda}\right)\,W_{-\lambda\bar{\epsilon},\frac{1}{4}}\,\left(
\frac{\epsilon\eta}{4\lambda}\right)+ c.c\right].\nonumber
\end{eqnarray}
Here $W_{\kappa,\mu}(z)$ is the Whittaker function \cite{GR};
$\epsilon=e^{i\pi/4}$, $\bar{\epsilon}=e^{-i\pi/4}$, and
$C(\lambda)$ is a real function yet to be determined.

Before we proceed with determining this function it is important to
establish its properties as $\lambda\rightarrow 0$. To this end we
note that the integral over $u$ in Eq.(\ref{moment1}) with $m=1$
must be divergent at small $u$. Indeed, if it is convergent then the
factor $\frac{1}{(m-2)!}=\frac{1}{\Gamma(m-1)}$ makes the first
moment equal to zero which contradicts the normalizability of wave
function $\int dx\langle |\psi(x)|^{2}\rangle=1$. At the same time
for $m>1$ the integral should converge to ensure finite higher
moments. This implies   that $\Phi(u,\phi)$ must tend to a constant
as $u\rightarrow 0$ \cite{rem3}. Given the asymptotic behavior of
Whittaker functions this is equivalent to:
\begin{equation}
\label{tile-non}
C(\lambda)=\lambda^{-\frac{3}{2}}\;\tilde{C}(\lambda),\;\;\;\tilde{C}(0)={\rm
const}.
\end{equation}
GF defined by Eq.(\ref{canon}) is periodic in $\phi$ with the period
$\frac{\pi}{2}$ as it should be for $q=2$. This is guaranteed by the
adding of the {\it c.c} term in Eq.(\ref{canon}). What is not
automatically guaranteed is that $\Phi(\xi,\eta)$ is {\it smooth} as
a function of $\phi$ at $\phi=0$. We will see that it is the
requirement of {\it smoothness}  at $\phi=0$ which fixes (up to a
constant factor) the unknown function $\tilde{C}(\lambda)$.

Indeed, the discontinuity of derivatives at $\phi=0$ may arise from
the branching of the expression in Eq.(\ref{canon})  at a small
$\eta$. From the representation of the Whittaker function in terms
of the hypergeometric functions one concludes that the general
solution Eq.(\ref{canon}) is a sum of a part which is regular in the
vicinity of $\eta=0$ and a part which has a square-root singularity
$\sqrt{\eta}\approx \sqrt{u} |\phi|$. The condition that this latter
part cancels out in the solution Eq.(\ref{canon}) is the following
($t$ is real):
\begin{eqnarray}
\label{rot} &&\Im
\left[\frac{\tilde{C}(\bar{\epsilon}t)}{\Gamma\left(\frac{1}{4}-it\right)}\,
e^{-\frac{i\eta}{8t}}\,_{1}F_{1}\left(
\frac{3}{4}-it,\frac{3}{2},\frac{i\eta}{4t}\right)\right]=0.
\end{eqnarray}
The crucial fact for the possibility to fulfil this condition is the
identity for the hypergeometric functions \cite{GR}:
\begin{equation}
\label{gr-iden-part}
e^{-z/2}\,_{1}F_{1}\left(\frac{3}{4}-it,\frac{3}{2},z
\right)=e^{z/2}\,_{1}F_{1}\left(\frac{3}{4}+it,\frac{3}{2},-z
\right).
\end{equation}
Now one can immediately guess the solution for $\tilde{C}(\lambda)$:
\begin{equation}
\label{GG} C_{0}(\lambda)=\Gamma\left(\frac{1}{4}+\epsilon\lambda
\right)\,\Gamma\left(\frac{1}{4}+\bar{\epsilon}\lambda \right).
\end{equation}

It is easy to see that the general solution to Eq.(\ref{rot})is
\begin{equation}
\label{gen-C}
\tilde{C}(\lambda)=C_{0}(\lambda)S(\lambda)=C_{0}(\lambda)\,\sum_{k=0}^{\infty}a_{k}\,\lambda^{4k},
\end{equation}
where the function $S(\lambda)$ must be regular in the entire
complex plane of $\lambda$. Now we apply the condition of
convergence of the integral over $\lambda$ in Eq.(\ref{canon}) at
large $\lambda$ to find the allowed asymptotic behavior of
$S(\lambda)$ at $\lambda\rightarrow\infty$. Substituting
Eq.(\ref{gen-C}) into Eq.(\ref{canon}) and using the asymptotics of
the Whittaker and $\Gamma$-functions we find that the integrand
behaves as $\lambda^{-3}S(\lambda)$ at $\lambda\rightarrow\infty$.
This means that $|S(\lambda)|$ should increase not faster than
$\lambda^{2}$. There is only one such entire function with the
structure of Eq.(\ref{gen-C}): this is a constant
$S(\lambda)=a_{0}={\rm const}$.

--{\it Conclusion and discussion.}
Eqs.(\ref{canon}),(\ref{tile-non}),(\ref{GG}) is the main result of
the paper. They give an exact and unique (up to a constant factor)
solution for the generating function at  $f=\frac{1}{2}$ anomaly
which can be used to compute all local statistics of the
one-dimensional Anderson model  at $L\rightarrow\infty$. The
integrability of TME Eq.(\ref{TME}) suggests that there is a hidden
symmetry of the problem at $f=\frac{1}{2}$. We make a conjecture
that this symmetry is naturally formulated in the three dimensional
space rather than in the two-dimensional space $(\xi,\eta)$ and that
it has to do with the symmetry of the 3d harmonic oscillator. This
conjecture is based on an analogy between our main result
Eq.(\ref{canon}) and the expression for the Green's function of the
3d harmonic oscillator problem \cite{BV}. This analogy concerns the
parameter ($\lambda$ in our problem and $k$ in Ref.\cite{BV})
entering both in the argument of the Whittaker functions and in its
first index in a mutually reciprocal way, as well as the second
index of the Whittaker functions being $\frac{1}{4}$ in both cases.
Establishing this symmetry would also be useful for studying the
anomalies at $f=\frac{p}{q}$ with $q>2$. We have obtained
\cite{rem4} the operator $\hat{L}_{f}^{(1)}(u,\phi)$ in
Eq.(\ref{expan}) and shown that the mechanism similar to
Eq.(\ref{id1}) is responsible for the anomaly at $f=\frac{1}{3}$.
The results of this study  will be published elsewhere.

 --{\it Acknowledgement.} We appreciate stimulating discussions with A.Agrachev,
Y.V.Fyodorov, A.Kamenev and A.Ossipov and a support from RFBR grant
06-02-16744.


\begin{thebibliography}{99}
\bibitem{rem1} All the positive moments of this distribution are
finite, in particular $\langle |\psi|^{2} \rangle=\frac{1}{L}$.
\bibitem{rem2} Note that the "Hamiltonian" in Eq.(\ref{H-1d}) is a
non-Hermitean operator. This is a consequence of the singular
inverse-quare potential. To make it Hermitean one has to impose a
condition $\varphi(0)=0$ assuming a hard wall at $\xi<0$. There is
no such condition in our problem.
\end{thebibliography}
\end{document}